# A Global South Strategy for Evaluating Research Value with ChatGPT

Robin Nunkoo
Department of Management, University of Mauritius, Reduit MU 80837 Mauritius
School of Tourism and Hospitality, University of Johannesburg, South Africa
College of Hotel & Tourism Management, Kyung Hee University, Seoul, Republic of Korea
School of Hospitality, Tourism & Events; Centre for Research and Innovation in Tourism (CRiT), Taylor's University, Malaysia.
Mike Thelwall
School of Information, Journalism and Communication, University of Sheffield, UK.

Research evaluation is important for appointments, promotions, departmental assessments, and national science strategy monitoring. Whilst Global North universities often have sufficient senior researchers for effective peer review and enough trust in citation data to use it for supporting indicators, the same is less likely to be true in the Global South. Moreover, Global South research priorities may not align well with citation-based indicators. This article introduces a ChatGPT-based strategy designed to address both limitations, applying it to Mauritius. The strategy involves giving ChatGPT instructions about how to evaluate the quality of research from the perspective of a given Global South nation and then using it to score articles based on these criteria. Results from Mauritius show that ChatGPT's scores for 1,566 journal articles published between 2015 and 2021 have an almost zero correlation with both ChatGPT research quality scores and citation rates. A word association thematic analysis of articles with relatively high scores for value to Mauritius identified a range of plausible themes, including education, policy relevance, and industrial production. Higher scoring articles also tended to mention the country or an important commercial sector in the abstract. Whilst the evidence suggests that assessing the direct value to a country of journal articles using ChatGPT gives plausible results, this approach should be used cautiously because it has unknown accuracy and ignores the wider value of research contributions.

## Statements and declarations

The research is funded by the Economic and Social Research Council (ESRC), UK (APP43146).

## Introduction

Expert review is often regarded as the best method to evaluate the quality of academic research for jobs and for departmental evaluations (Wilsdon et al., 2025, Hicks et al., 2015), but it is time consuming and dependant on the availability of relevant expertise. Citation-based indicators are sometimes used instead if expertise is unavailable or judged too costly (Sivertsen, 2017). For example, Malaysia's indicators-based system (Ta et al., 2021) is likely to be much cheaper than the UK's expert review. The two approaches are frequently also employed in combination, with indicators supporting expert judgement (Moed, 2005). From a Global South perspective, all three strategies may be problematic. In a developing country with an expanding research system, a much smaller per-capita research capacity, and lower life expectancy, there may be too few senior researchers to effectively evaluate research for most purposes. Citation analysis is also weaker in the Global South because bibliometric databases tend to focus on English-language publications and the Global North in general (Koch & Vanderstraeten, 2021; Mongeon & Paul-Hus, 2016; Vera-Baceta et al., 2019), so citation counts predominantly reflect Global North interests and priorities. More fundamentally, citation-based indictors may poorly reflect the needs of Global South nations, for example because they most directly reflect contributions to global scientific knowledge whereas Global South priorities may include technology transfer, capacity building, and discoveries with local relevance (Barrere, 2020).

Although previous scholars have identified the above problems for Global South research evaluation, no solutions seem to have been adopted. The closest relevant strategy is perhaps to classify articles for their Sustainable Development Goals (Hajikhani & Suominen, 2022) but this does not focus on the needs of an individual country and a Global South nation can reasonably hope to extract direct use from its researchers. A potential solution is to use large language models (LLMs) instead of human experts or citation-based indicators on the basis that LLMs avoid the need for expert evaluations and may be adaptable, at least in theory, to Global South needs. Previous research has shown that ChatGPT 4o, ChatGPT 4o-mini and Google Gemini 1.5 Flash can score academic publications for a standard Global North definition of research quality based on rigour, significance and originality, and that the ChatGPT scores correlate more highly with human expert quality scores than do some citation-based indicators in most fields (Thelwall, 2025b). Thus, it is logical to assess whether this strategy could be adapted to a Global South research quality definition, bypassing both experts and citations. This might, in theory, give cheap and targeted research evaluation data.

In response to the above, this article introduces a strategy to harness LLMs for Global South research evaluations, focusing on journal articles as the main scholarly output in most fields. Evaluating this strategy is difficult because all countries are different, and no country has published a "gold standard" set of articles with validated research quality scores. For this reason, a case study of a single country, Mauritius, is presented and the research questions focus on evaluating plausibility rather than accuracy. ChatGPT was chosen ahead of other LLMs because previous comparisons have shown that it performs better at research quality scoring than Gemini, and informal testing suggests that it is also better than DeepSeek for this task (Thelwall, 2024, 2025ab). The country Mauritius was chosen for the case study for pragmatic reasons: the second author is from the country and has devised its current national research strategy after extensive consultations with the Honorable Minister of Tertiary Education, Science and Research and relevant stakeholders. He therefore is well placed to create guidelines for a national research quality framework for ChatGPT and to evaluate the results. Mauritius is a developing small island state in the Indian Ocean in Southern Africa. It has a population of a million serviced by five public universities and many private teaching-focused higher educational institutions, including some based in the UK and Australia. It is historically successful, with rising standards of living and reducing inequality since independence from the UK in 1968.

The following research questions drive the study. Two strategies are tested: evaluating research quality in a way that is oriented towards value to Mauritius and directly assessing value to Mauritius.

- RQ1: Do Mauritius-oriented ChatGPT research quality scores give different values to ChatGPT scores for research quality and to citation rates?
- RQ2: Do ChatGPT scores for value to Mauritius give different values to ChatGPT scores for research quality and to citation rates?
- RQ3: Which types of research gets different scores for value to Mauritius than for research quality?

## Methods

The research design was to create prompts for ChatGPT that evaluate (RQ1) research quality from the perspective of Maritius, and (RQ2) usefulness to Mauritius, and then to assess the extent to which the scores correlate with each other and general research quality scores for articles from Mauritius, together with citations. The next stage (RQ3) was to identify the topics that score high or low on each of the three criteria relative to the others, using these to inform a discussion of whether the scores plausibly evaluate the targeted type of research quality. The following is the broad outline of the research design.

1) Download all journal articles with an affiliation country of Mauritius from Scopus, published after 2014 (to be not older than ten-year age cutoff to keep the data relatively current) and before 2022 (to allow over three years of citations for a robust citation analysis).
2) Remove articles without abstracts or with very short abstracts (since the scores are based on abstracts, and short abstracts often indicate short articles or contributions that are not full articles). The shortest 10% were removed.
3) Split the articles into four groups to match the UK REF quality definitions previously used for ChatGPT research evaluation trials.
    a. A: Health and life sciences
    b. B: Physical sciences, maths and engineering
    c. C: Social sciences
    d. D: Arts and humanities
4) Make three different system instructions for ChatGPT, describing how to evaluate the articles (examples attached).
    a. Standard research quality instructions (originality, significance, rigour), taken from the UK REF2021.
    b. Standard research quality instructions as above but modified to state that the significance must be just for Mauritius.
    c. New instructions asking for a score indicating value to Mauritius. This is essentially removing originality and rigour from (b), but also with different names for the scoring levels.
5) Submit the articles from 3) above to ChatGPT separately with each of the three system instructions from 4) above.
6) Analyse the results to address the research questions (see analysis section).

### *Data: Articles from Mauritius*

The dataset consisted of all standard journal articles in Scopus published 2015-2021 with any author having an affiliation from Mauritius.. Journal articles were chosen since these are the standard research outputs in most fields and are more systematically collected than other output types by bibliometric databases. Only standard journal articles were included since other types, such as editorials and reviews, tend not to contain primary research. No requirement was made for a first (or corresponding) author to be from the country because research evaluations often (e.g., the UK REF) treat all authors as equal. The year range was chosen to include relatively recent research, since national priorities change over time.

The articles were downloaded from Scopus using its web interface with the query below on 5 July 2025. Articles published before 2015 or after 2021 were filtered out. After downloading, articles without abstracts or in the shortest 10% of abstracts were removed.
AFFILCOUNTRY(Mauritius) AND (LIMIT-TO(DOCTYPE, "ar")) AND ( LIMIT-TO( LANGUAGE, "English"))

### *ChatGPT scores (standard and probability approaches)*

ChatGPT can be asked to score an article for research quality by configuring it with system instructions defining a quality scoring system and then entering the article title and abstract as the user prompt, with a request to "Score this journal article". Previous research has shown that ChatGPT 4o and 4o-mini outperform Google Gemini 1.5 Flash for this (Thelwall, 2025ab) and, surprisingly, that optimal scores are obtained when the article title and abstract is entered but not the full text (Thelwall, 2024). This suggests that ChatGPT does not have a genuine ability to evaluate research but only guesses at an appropriate score, such as by cross-referencing the quality definition with summary information from the article, perhaps mediating its decision with some wider knowledge of research or society. As for citation-based indicators, the positive correlations found for ChatGPT scores can still serve a useful purpose to support research evaluations despite not being genuine evaluations.

For this study, only article titles and abstracts were used as inputs, for the highest accuracy, and three sets of system instructions were employed, as defined below. They were submitted to ChatGPT 4o-mini using its Applications Programming Interface (API) during July 2025 five times. Using the API allows system instructions to be fully exploited and the queries to be automated. Each query was submitted five times in separate sessions and the scores extracted and averaged. Averaging the scores substantially improves the value of the results (their correlations with expert judgements), apparently because they give insights into ChatGPT's confidence in its scores (Thelwall, 2025ab).

For each prompt, the above approach gives a ChatGPT report containing a score and a rationale for the score, based on the system instructions (see below). An alternative approach that gives slightly more accurate results is to request just a score without the associated report and to replace the recommended score with a probability-weighted average score from the token probabilities returned by ChatGPT (Thelwall & Yang, 2025). For example, if ChatGPT reports a score of 3* but its token probabilities reveal that the 3* score had a 0.6 probability and that 2* had a 0.4 probability then the article would be assigned a weighted average score of 0.6*3+0.4*2 = 2.6*. The probability approach was also used five times for each article and set of system instructions and compared with the standard approach to cross-validate both and then combined for the most accurate results overall. Although ten iterations of the probability approach would probably have given more accurate results, the reports from the non-probability method are useful to analyse the results and using two approaches allows cross-validation.

**ChatGPT research quality (Global North) system instructions**

Most previous research evaluations with ChatGPT have configured it with research quality guidelines from the UK REF2021 research evaluation system to score articles (Thelwall, 2025ab). These guidelines define the three commonly used quality criteria of originality, significance and rigour (Aksnes et al. 2019; Langfeldt et al., 2020). Prior studies have slightly adapted them as system instructions for ChatGPT in four flavours: Health and life sciences (REF Main Panel A), physical sciences and engineering (REF Main Panel B), social sciences (REF Main Panel C), and arts and humanities (REF Main Panel D). These are implicitly Global North criteria since they originate from a Global North country and the criteria are not unusual for the region. They emphasise general dimensions of quality without any country-specific factors, although the practical value of research is acknowledged in the significance dimension.

For the current paper, these system instructions were re-used as the Global North quality criteria representative. They have been previously published (Thelwall, 2025b), and are republished in the online appendix (https://doi.org/10.6084/m9.figshare.29730257) for the current paper.

**ChatGPT Mauritius-oriented research quality system instructions**

There is no Global South research quality definition. Whilst Global North goals have been challenged as irrelevant (Barrere, 2020), there does not seem to be an agreed Global South set of criteria (e.g., to replace originality, significance, rigour), and there is not yet an agreed research quality approach for Mauritius. Thus, a new definition must be proposed here and converted into ChatGPT instructions. For this, we take an extreme view to focus on the primary direct value to a developing nation, ignoring the wider benefits (e.g., university reputation, research capacity) of having researchers producing studies that would be regarded as high quality in the Global North.

A set of system instructions for ChatGPT was created for the needs of Mauritius by modifying the generic research quality instructions from the UK REF (see Appendix). Whilst some benefits are relatively generic, such as for health (although key health needs vary by country), countries vary substantially in their major economic sectors. Thus, the instructions were modified to specify the major economic activity sectors in Mauritius. These were taken primarily from Table 4 (Gross Value Added by industry group) of the 2024 annual report of the Bank of Mauritius (2024). In addition, knowledge of the relevance of academic research to the economy in Mauritius is part of the expertise of the first author, who is designing the national research evaluation strategy for the Ministry of Tertiary Education, Science and Research in Mauritius at the time of writing.

**ChatGPT Value to Mauritius system instructions**

A set of system instructions for direct value to Mauritius was created by modifying the second set to remove the rigour and originality requirements so that they just focus on assessing direct significance to Mauritius (see Appendix). The quality definitions were also modified to align them with significance for Mauritius. For this, it would not be coherent to have "world-leading" or even "Quality" ratings, as in the standard definitions:
- 4*: Quality that is world-leading in terms of originality, significance and rigour.
- 3*: Quality that is internationally excellent in terms of originality, significance and rigour but which falls short of the highest standards of excellence.

- 2*: Quality that is recognised internationally in terms of originality, significance and rigour.
- 1*: Quality that is recognised nationally in terms of originality, significance and rigour.

Thus, the definitions were changed to the following for the significance-only system instructions. This targets Mauritius and avoids mentioning quality or international standards.
- 4*: Research that makes an exceptionally valuable contribution to Mauritius.
- 3*: Research that makes a very valuable contribution to Mauritius.
- 2*: Research that makes a valuable contribution to Mauritius.
- 1*: Research that makes a minor contribution to Mauritius.

**Field classification of articles**

There were four different versions of the original system instructions, with each customised for one broad area of research, as covered by the REF. They are Main Panel A (health and life sciences and psychology), B (physical sciences, maths and engineering), C (social sciences) and D (arts and humanities). The above process therefore produced four corresponding versions of each system instructions (A, B, C and D).

The articles from Mauritius were assigned to one of A, B, C and D as follows. Each article has one or more broad field classifications in Scopus, which is derived from its journal's All Science Journal Classification (ASJC: https://supportcontent.elsevier.com/RightNow%20Next%20Gen/SciVal/ASJC1.xlsx) codes, and articles were assigned to a field based on these as follows:
- A: 11 (Agricultural and Biological Sciences), 13 (Biochemistry, Genetics and Molecular Biology), 24 (Immunology and Microbiology), 27 (Medicine), 28 (Neuroscience), 29 (Nursing), 30 (Pharmacology, Toxicology and Pharmaceutics), 32 (Psychology), 34 (Veterinary), 35 (Dentistry), 36 (Health Professions)
- B: 15 (Chemical Engineering), 16, 17, 18 (Decision Sciences), 19 (Earth and Planetary Sciences), 21 (Energy), 22 (Engineering), 23 (Environmental Science), 25 (Materials Science), 26 (Mathematics), 31 (Physics and Astronomy)
- C: 14 (Business, Management and Accounting), 20 (Economics, Econometrics and Finance), 33 (Social Sciences)
- D: 12 (Arts and Humanities)

When articles had classifications matching multiple fields, they were manually assigned to one of the four categories based on their publishing journal, title and abstract.

**Citation scores**

The citation counts from Scopus for the articles cannot be directly compared with the ChatGPT scores because older articles have the unfair advantage of a longer time to attract citations. This problem can be solved by replacing citation counts with a field and year normalised citation indicator, for which each citation count is divided by the average for its field and year. After this calculation it is fair to compare between fields and years.

For this article the Normalised Log-transformed Citation Score (NLCS) field and year normalised indicator was used (Thelwall, 2017), with the fields being the four panels. Narrower fields are usually used, but the field classification of Scopus is not very useful for this purpose because it is journal-based and can assign journals to fields even when there is only a partial match, so the panel classifications described above were used instead. A log-normalised indicator was used to reduce the impact of skewing on the results. This is important because otherwise individual highly cited articles can have a large influence on the results for other articles.

For the NLCS, each citation count c was replaced by ln(1+c), where the 1 offset is necessary because the log of 0 is undefined. Then, each ln(1+c) was divided by the mean of the ln(1+c) values for all articles in the dataset from the same panel and year. After this, the mean NLCS for each main panel and year is 1 and it is fair to combine these values from different years for correlation tests.

### *Analysis*

For RQ1 and RQ2, all sets of scores were correlated against each other to identify the extent to which they give different results. Spearman correlations were used because the rank order of scores from ChatGPT is more important than their absolute values (Thelwall & Yang, 2025). The correlations were calculated separately for each Main Panel to reduce the influence of disciplinary differences. Bootstrapped confidence intervals were calculated because correlation coefficients are imprecise at the sample sizes analysed here.

Two methods were developed to address RQ3 (Which types of research gets different scores for value to Mauritius than for research quality). Both are approaches to compare articles that have the biggest differences in scores for value to Mauritius and research quality. First, for each article, its value to Mauritius score was subtracted from its research quality score. For this indicator difference quantity, high values associate with a relatively high research quality and low values with a relatively high value to Mauritius.

a. The 50 highest and 50 lowest scoring articles on the indicator difference indicator for each Main Panel (10 for Main Panel D, with only 21 articles) were examined to identify whether one set was plausibly more oriented towards benefits to Mauritius than the other. This was achieved by searching the article titles and abstracts for a mention of the country, backed by a manual check of anomalies. This is a simplistic method but has the advantage of transparency. The sample size of 50 was chosen in advance to give a set of articles that was not too large to manually check individually (320 in total) although this did not prove necessary.
b. A Word Association Thematic Analysis (WATA) (Thelwall, 2021) was conducted to identify themes in the articles with relatively high and low indicator difference values, based on terms in their titles and abstracts. WATA works by first identifying terms that occur statistically significantly more frequently in one set of texts (here the higher scoring half of the articles) than the other, using a chi squared test, backed by a Benjamini-Hochberg (Benjamini & Hochberg, 1995) familywise error rate protection procedure. These words were then manually assigned a context or contexts by reading at least ten random articles containing them (e.g., "Fruit" was assigned the context, "agriculture, fly, consumption"). The words with their contexts were then manually clustered into themes for reporting. This is a more complex method than the first one and is partly subjective but can give finer grained insights into the types of articles that are higher and lower scoring.

# Results

The mean scores for the different prompts tend to be lowest for the value for Mauritius prompt, with Panel A (health and life sciences) articles tending to get the highest scores for research quality. In contrast, panel C (social sciences) scores the highest for value to Mauritius (Table 1).

Table 1. Mean values for the three types of ChatGPT score in each of the four panels (A: Health and life sciences; B: Physical sciences and engineering; C: Social sciences; D: Arts and humanities). Data: 1566 journal articles in Scopus without short abstracts published 2015-2021 with at least one author with a Mauritius affiliation.

| Score\Main Panel | A | B | C | D |
|---|---|---|---|---|
| Quality | 2.97 | 2.68 | 2.64 | 2.82 |
| Quality (Mauritius) | 3 | 2.75 | 2.72 | 2.88 |
| Value for Mauritius | 2.26 | 1.86 | 2.59 | 1.82 |
| **Articles** | **791** | **447** | **307** | **21** |

Scores from standard prompts tend to correlate highly with scores from the probability prompts (Figure 1). The extremely high correlations for value to Mauritius perhaps reflect a simpler task in the sense of a single concrete dimension rather than three more abstract dimensions.

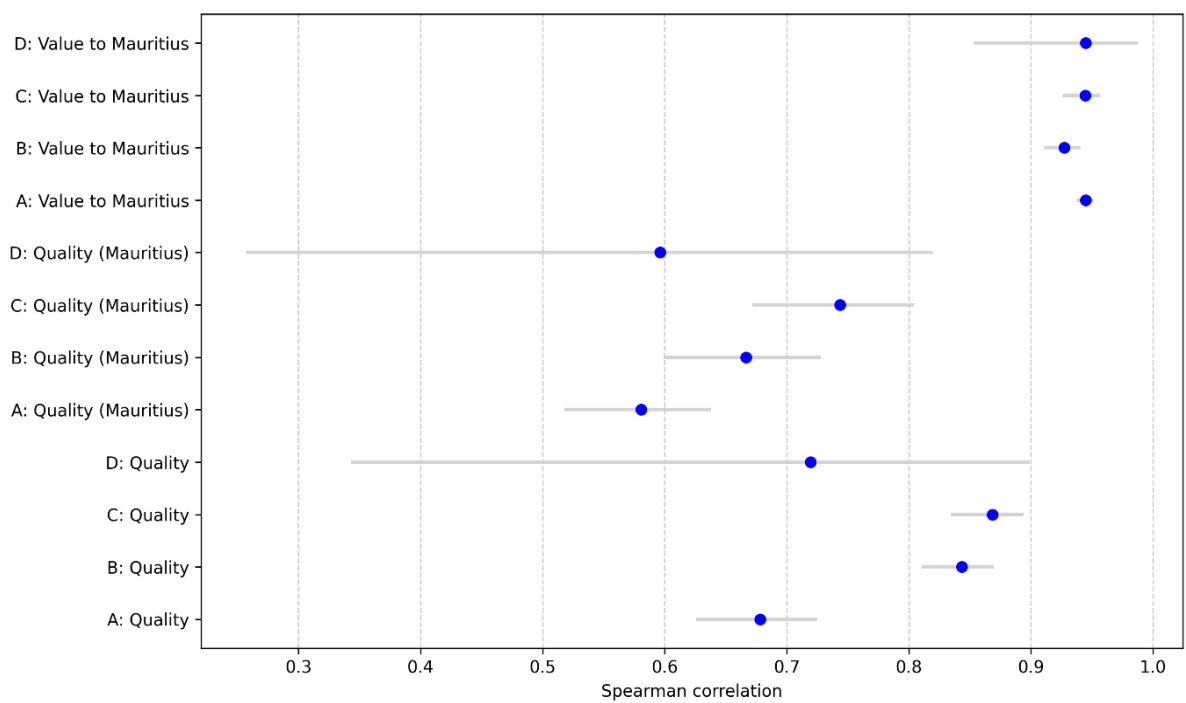

Figure 1. Spearman correlation between ChatGPT scores from standard prompts (average of five) and probability prompts (average of five) by Main Panel and score type. Error bars show bootstrapped 95% confidence intervals.

## *RQ1,2: Research quality vs. value to Mauritius vs. citations*

The correlations between research quality scores and value to Mauritius scores are very weak and either negative (C: social sciences) or with confidence intervals including 0.

Thus, there seems to be no relationship between research quality and value to Mauritius for research (Figure 2).

There are very strong correlations between research quality and Mauritius-oriented research quality, perhaps because two of the three dimensions are the same (rigour and originality) and the third has some overlap (significance vs. significance to Mauritius). The weak to moderate positive correlations between value to Mauritius and Mauritius-oriented research quality presumably reflect a partial overlap in only one of the three dimensions of research quality (Figure 2).

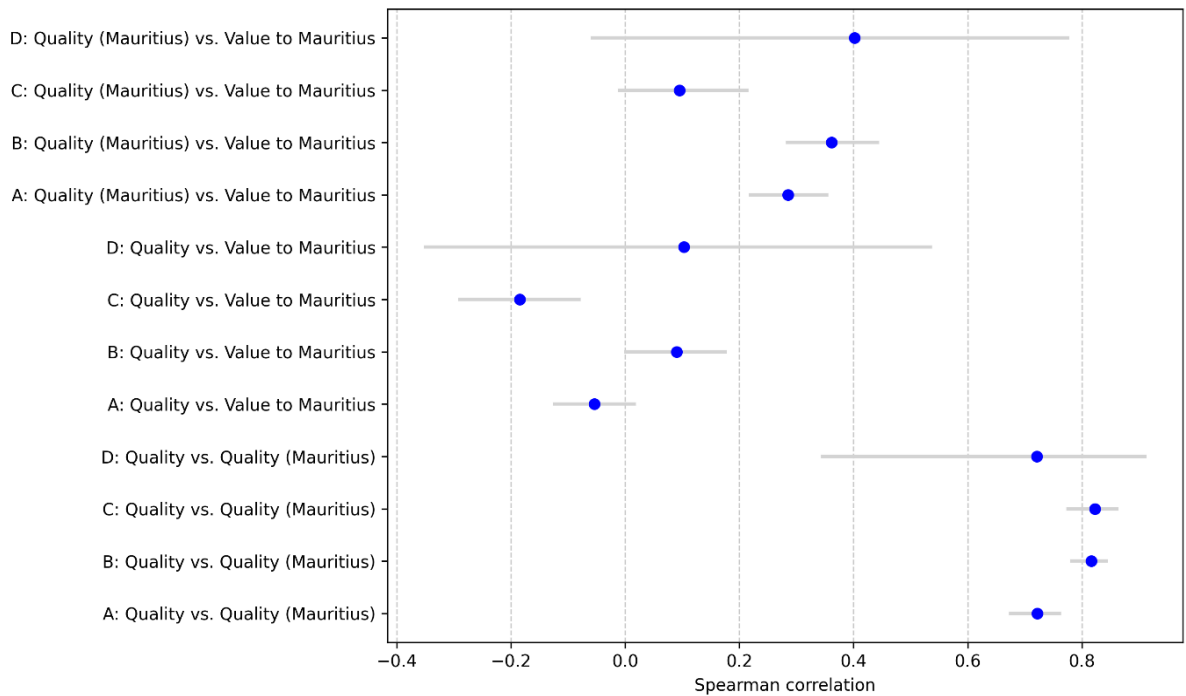

Figure 2. Pairwise Spearman correlations between ChatGPT scores from the three different types of prompts (average of ten, averaging standard and probability prompt scores) by Main Panel. Error bars show bootstrapped 95% confidence intervals.

The citation-based indicator has mostly weak negative correlations with value to Mauritius but a weak to moderate positive correlation with research quality. It is perhaps surprising that there are not stronger negative correlations between citations and value to Mauritius given that Mauritius is a small developing country so there are few researchers to cite work for its value to Mauritius. The lack of a stronger negative correlation may reflect many contributions having general value, not just to Mauritius.

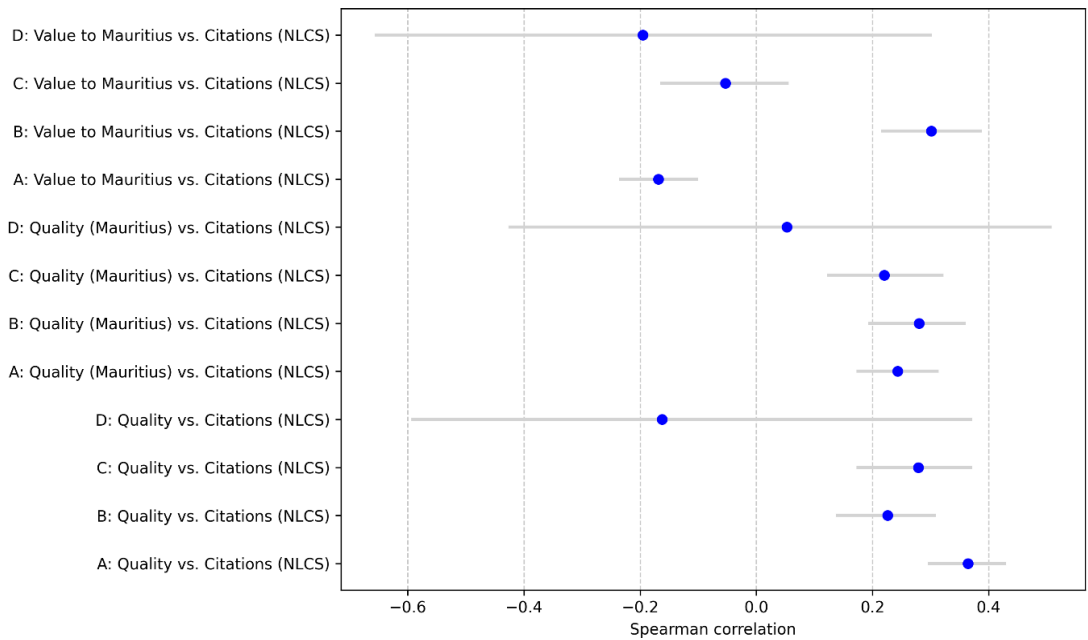

Figure 3. Spearman correlations between ChatGPT scores from the three different types of prompts (average of ten, averaging standard and probability prompt scores) and citation rates (NLCS) by Main Panel. Error bars show bootstrapped 95% confidence intervals.

### *RQ3a: Article types with relatively high or low value to Mauritius*

Within each of the four panels, the articles were sorted by their value to Mauritius score subtract their quality score for a qualitative investigation of the types of articles with a relatively high and low value to Mauritius. The most obvious pattern was that the top articles tended to mention Mauritius and the bottom articles tended not to. This can be illustrated by examining the top and bottom 50 for each Main Panel.

- A: 48 articles mentioned Mauritius in the top 50 compared to 1 in the bottom 50.
- B: 47 articles mentioned Mauritius in the top 50 compared to 0 in the bottom 50.
- C: 50 articles mentioned Mauritius in the top 50 compared to 0 in the bottom 50.
- D: 9 articles mentioned Mauritius in the top 10 compared to 1 in the bottom 10 (out of 21).

Examining the exceptions in the large datasets (A, B, C) gives additional insights. Each article has five ChatGPT reports but the comments below are all taken from the first one to reduce the risk of bias through cherry-picking.

- **High** value to Mauritius article not mentioning Mauritius (panel A). "A pre and post survey to determine effectiveness of a dietitian-based nutrition education strategy on fruit and vegetable intake and energy intake among adults" involved a nutrition intervention in Mauritius, but this was not mentioned in the title or abstract. The article was open access from 2016, so ChatGPT may have detected the Mauritius context. One ChatGPT report initially claimed that fruit intake was relevant to Mauritius, without stating that the intervention was in Mauritius, "Given the implications for health improvement, especially among the adult population, this research offers substantial insights into enhancing nutritional behaviors, which are vital for reducing health-related issues in Mauritius." At the end of the report, it seemed to suggest that the intervention was in Mauritius

- without directly stating it, "Additionally, exploring collaborative efforts between nutritionists and local chefs to create appealing vegetable dishes could further aid in overcoming taste barriers, thus broadening the impact of such educational interventions in Mauritius. Overall, the study represents a solid contribution to understanding and enhancing dietary practices within the country, warranting a high score of 3*." Thus, it is possible that a combination of imperfectly remembered information about the article online, and the importance of fruit consumption to Mauritius has given its high score.
- **High** value to Mauritius article not mentioning Mauritius (panel A). "Development of an Irrigation Scheduling Software for Sugarcane" again is highly relevant to Mauritius (for the sugarcane industry) but does not mention the country. The full text mentions Mauritius as one of four countries where the software was tested. The first report does not mention where the research was conducted but mentions the importance of the sugarcane industry to Mauritius, "The developed irrigation scheduling software addresses a critical issue in the sugarcane sector of Mauritius, where efficient water management is paramount due to limited rainfall."
- **Low** value to Mauritius article mentioning Mauritius (panel A). "A new telemetry-based system for assessing cardiovascular function in group-housed large animals. Taking the 3Rs to a new level with the evaluation of remote measurement via cloud data transmission" mentioned Mauritius as a location where some Cynomolgus monkeys are kept, one of the two species that the new system was tested on. Nevertheless, the first report concludes that, "the niche nature of the application, along with limited immediate relevance to broader Mauritian sectors such as human healthcare and economy, justify a score of 2*."
- **High** value to Mauritius article not mentioning Mauritius (panel B). "Sustainable concrete: Potency of sugarcane bagasse ash as a cementitious material in the construction industry". The first ChatGPT report emphasises value to multiple sectors in Mauritius of the approach, "The research addresses the utilization of sugar-cane bagasse, a significant agricultural by-product in Mauritius. Given that Mauritius primarily relies on sugarcane as a key sector of its economy, this work has the potential to not only reduce waste but also add value to an existing agricultural commodity. By enhancing the performance of polylactide and polydioxanone electrospun scaffolds for tissue engineering, it opens avenues in both biomedical and economic sectors, notably in biotechnology and health sectors, which are crucial for Mauritius' development."
- **High** value to Mauritius article not mentioning Mauritius (panel B). "Production of cardboard from waste rice husk". This article mentions Mauritius twice in its conclusion as an application area, "Use of non-wood plant fibres for pulp production in Mauritius is not a matter of choice but an obligation. Mauritius is not rich in wood resources" but not on the title or abstract. The first ChatGPT report justifies a high score on the basis that the problem is relevant to Mauritius, "The study addresses a significant environmental issue related to agricultural waste management in Mauritius, particularly concerning rice husk, which is a byproduct of rice cultivation - a common agricultural practice in the country."
- **High** value to Mauritius article not mentioning Mauritius (panel B). "Thermo-economics analysis of a cane bio-refinery". This article mentions Mauritius many times in the full text, which is available online, but not it its title and abstract. The first report justifies a high score because of the importance of the technology to Mauritius, "Considering that sugar cane is a significant agricultural product for Mauritius, influencing both its economy and agricultural practices, the findings of this study have the potential to promote significant operational improvements in existing facilities."

Thus, ChatGPT seems to be able to correctly infer importance to Mauritius for some contexts where the country is not mentioned in the article title/abstract fed to it. Two concern sugarcane, a high-profile product of Mauritius, but two others are less obviously associated with the country, fruit intake and rice husk waste. This suggests that ChatGPT is able to use information about Mauritius from its knowledge base to support its answers.

### *RQ3b: Themes for research with high and low value to Mauritius compared to research quality*

The WATA found thirteen substantive themes for articles that had relatively high value to Mauritius, compared to research quality, as well style and structured abstract themes (presumably encouraging value to be explicitly stated) (Table 2). The substantive themes primarily concern the type of value to Mauritius or the inclusion of Mauritius within the study. The two exceptions, interviews and surveys, seem to relate to studies collecting data from Mauritius, which would explain their value to the country.

Table 2. Themes identified from a word association thematic analysis of terms that occurred statistically significantly more frequently in the titles and abstracts of articles that had an **above median score for Mauritian value** compared to (i.e., subtract) quality.

| Theme* | Terms in theme |
|---|---|
| Development (economic, sustainable) | development |
| Education (university) | education |
| Food (diet, sources) | food |
| Fruit (agriculture, fly, consumption) | fruit |
| Impact/value | impact, need, potential |
| Interviews | interviews |
| Management (business/diabetes/disease) | management |
| Mauritius | island, Mauritian, Mauritius, small |
| Multiple/style (Sectors of the economy) | sector |
| Policy (economic, government) | government, policies |
| Production (industry, agriculture) | production |
| Solar energy | electricity, solar |
| Survey (of Mauritius) | survey |
| Structured abstract* | design/methodology/approach, findings, originality/value, purpose, research |
| Style* | be, conducted, study, was |
| Unclear (multiple unrelated contexts)* | main |

*Not a substantive theme.

The following quoted extracts from ChatGPT reports on value to Mauritius illustrate some of the above themes. In each case there is a direct claim of value to Mauritius. These claims can be cross-checked with the list of key sectors derived for the system instructions ("health, the economy, culture, or policy in Mauritius, and particularly for food processing, sugar cane, mining and quarrying, agriculture, agro processing, ocean economy, textiles and apparel, clothing, mineral, rum distilling and chemicals, metal products, renewable energy, transport equipment, medicine, nonelectrical machinery, tourism, biotechnology, agriculture, farming, financial, banking, insurance, construction, information and communication technology, real estate, public administration, defense, arts, entertainment and recreation, education, wholesale and retail trade, transportation and storage, manufacturing, arts and entertainment"), and comments based on matches with this list are added at the end of the quote in the list below.

- **Development**: "This research has significant implications for the energy policy landscape of the country - especially as Mauritius is moving towards renewable energy sources to diversify its energy mix and to comply with international commitments regarding sustainable development and climate change." Development is a core need for developing countries, so the theme seems uncontroversial, and the quote also mentions the key sector of renewable energy.
- **Education**: "Given that education is a cornerstone of Mauritian development, particularly in a resource-scarce environment, the issues raised about quality, financial implications, and regulatory bypassing are of paramount significance." Education is a key sector in the list above.
- **Food**: "This work is significant due to the critical role of honey bees in agricultural pollination, which directly impacts food security and agricultural productivity - a primary sector in Mauritius." Food is not mentioned directly but food processing, agriculture and farming are all key sectors of Mauritius.
- **Fruit**: "the effective larvicidal properties against Bactrocera zonata can reduce economic losses in the fruit sector, underscoring the work's relevance to Mauritius's agricultural economy." As above, fruit is not mentioned directly but food processing, agriculture and farming are all key sectors.
- **Mauritius/Management**: "The research addresses a critical environmental issue, which is the management of concrete waste in Mauritius." Management is not a key sector, but the term reflects a wide range of applications, as the quote illustrates. Furthermore, as a small island developing nation, Mauritius faces unique global and local sustainability challenges resulting from climate change and waste, respectively.
- **Policy**: "While the research does not yet show direct evidence of policy changes within Mauritius, its implications for similar regional mining contexts suggest potential policy adoption in the near future.", "The findings regarding soil erosion and its relationship to rainfall and vegetation dynamics are not only scientifically valuable but are also applicable to policy-making in Mauritius." Policy is mentioned as a key sector.
- **Production**: "The article presents a significant evaluation of bio-methane production potential from crop residues, which directly impacts the energy landscape in Mauritius." Production is not a key sector, is a near synonym of

- manufacturing, which is. This example also reflects Mauritius' emphasis on renewable natural gas from the decomposition of organic matter.
- **Sector**: "this article makes a very valuable contribution to understanding the economic implications of port sector investments and their social repercussions in Mauritius".
- **Solar energy**: "The research on solar water heater (SWH) efficiency is highly relevant to Mauritius, where renewable energy policies have been increasingly emphasized.", "The research provides crucial insights into the solar energy potential of Mauritius, a theme that resonates profoundly with Mauritius' national goals toward renewable energy development and sustainability." Renewable energy is a key sector.
- **Survey**: "The article explores the dynamics of multiculturalism in Mauritius, shedding light on how cultural diversity is perceived and managed within a secular state framework. [] The use of large-scale survey research among adolescents of diverse ethnic backgrounds [] adds empirical weight to the discussion." This is a method rather than a key sector.

There were seven themes from articles with a relatively low value to Mauritius compared to their research quality (Table 3). These seem to relate to more theoretical research, and taxonomic studies, as well as a theme focusing on another country (Turkey).

Table 3. Themes identified from a word association thematic analysis of terms that occurred statistically significantly more frequently in the titles and abstracts of articles that had an **above median score for research quality** compared to (i.e., subtract) Mauritian value.

| Theme | Terms |
|---|---|
| Chemical reactions | reaction |
| Computational approaches | computational, computed |
| Density Functional Theory (DFT) | density, DFT, electronic, functional, theory, vibrational |
| Extracts of plants from Turkey | Turkey |
| Novel entities | novel |
| Structures | structure |
| Taxonomy of plants and animals | asexual, described, descriptions, DNA, fungal, fungi, genera, genus, host, illustrations, introduced, LSU, molecular, morphological, morphology, new, notes, nov., phylogenetic, phylogeny, sequence, species, specimens, taxa, taxonomic, taxonomy, Thailand |
| Style* | based, propose, we |

*Not a substantive theme.

The following extracts from ChatGPT reports on value to Mauritius to illustrate some of the above themes. In each case a justification for a low score is based on a lack of evidence of direct benefits to Mauritius. ChatGPT seems to comment often that there is a lack of claimed value to Mauritius although in other reports it sometimes infers such a connection even if it is not explicitly stated in the title or abstract. None of these directly match one of the key sectors in the list generated for the system instructions, with the partial exception that computational approaches fall within the broad remit of information and communication technology.

- **Chemical reactions**: "This article primarily discusses the intricacies of a specific chemical reaction - the Fischer indole synthesis - focusing on theoretical insights into the influence of fluorine substituents. However, its application or influence on policy or practice in Mauritius, especially concerning local industries such as agriculture, agro-processing, or pharmaceuticals, appears to be minimal."
- **Computational approaches**: "The research presented in the article focuses primarily on the theoretical analysis of boradiazaindacene fluorophores using advanced computational methods. While the topic has scientific merit, its direct implications for various sectors in Mauritius are limited. The specific application of these compounds in industry (such as the textiles or pharmaceuticals sectors) is not sufficiently articulated within the context of Mauritius' economic landscape."
- **Density Functional Theory (DFT)**: "The research focuses on a specialized area within organic chemistry, specifically the [3+2] cycloaddition reaction involving phosphorinium compounds. While the study employs density functional theory (DFT) to explore the thermodynamic and kinetic parameters of the reaction, its direct relevance to Mauritius seems limited."
- **Extracts of plants from Turkey**: "The findings presented in this article largely focus on the phylogenetic classification and morphological details of newly identified saprobic fungal species from specific plant substrates in Turkey and Italy. While the research adds to the taxonomic knowledge of Ascomycetes and has scientific merit, it does not directly address or improve health, the economy, culture, or policy in Mauritius."
- **Novel entities**: "This research presents a novel compound and complex involving oxomolybdenum, which primarily holds potential interest within the fields of inorganic chemistry and materials science. While the article showcases robust synthetic and theoretical investigations that might have applications in specialized areas like catalysis or materials development, the direct applicability to the key sectors in Mauritius is limited."
- **Structures**: "The findings related to the structure and bonding of metal-cyclen complexes do not appear to have any direct implications for key sectors such as agriculture, tourism, or manufacturing in Mauritius."
- **Taxonomy of plants and animals**: "The research focuses specifically on a taxonomic and phylogenetic reappraisal of certain fungal genera []. Fungi play various ecological and economic roles, yet the study is so specialized that it limits its direct applicability to service users and researchers in Mauritius. Without a clear connection to local agricultural systems, biotechnological applications, or environmental policies, the study fails to establish a practical framework through which findings could benefit sectors crucial to Mauritius, such as agriculture or biotechnology."

# Discussion

This article is limited by several factors that could influence the conclusions. The results might be different for alternative prompts. This is especially true for the novel value to Mauritius prompts, which are the result of decisions about which information to include and how to name the scoring system. The results may also not generalise to other countries, especially if they are less specialised so that it is hard to delineate their main economic activity sectors. Other LLMs may also have given different results.

The results are not directly comparable to any previous studies since none have compared scores from different research quality or value definitions for a set of articles. A minor exception is that moderate correlations between research quality scores from ChatGPT and citations has been observed before, for the UK (Thelwall, 2025b).

## *Disciplinary differences*

There are some broad disciplinary differences in the results. The social sciences (panel C) are relatively strong on the value to Mauritius indicator compared to the other panels, A (health and life sciences) and B (physical sciences and engineering). This seems to be primarily due to social science research much more frequently having an explicitly Mauritian context. For example, the percentages of titles/abstracts mentioning Mauritius are: A 27%, B 23%, C 55%, D 48%.

The greater Mauritian focus of social science research with at least one author from the country is presumably due to social issues being more geographically diverse than health, physical sciences and engineering. Life science is also geographically diverse, however, at least at the organism level. Thus, social science research from Mauritius might tend to be more specifically relevant to Mauritius than most research from panels A and B. It seems likely that social science research needs to be more country specific to be useful than does physical sciences and engineering research, however. This may also explain why panel B (physical sciences and engineering) has the lowest value to Mauritius score other than for D, which has a too small sample size to be relevant here (n=21).

## *Is it reasonable to directly assess value to Mauritius?*

The results suggest that tailoring research quality definitions to Mauritius produces a small difference in the results but replacing them with value to Mauritius, as defined above, gives completely different results. The value to Mauritius results are oriented towards higher scores for articles explicitly stating value to Mauritius or importance to a recognised key sector of Mauritius in their titles or abstracts.

It is both a conceptual and policy issue to decide whether this orientation is desirable for any given research evaluation. Whist value to Mauritius scores focus on immediate benefits to the country, they ignore longer term and wider value.

From short-term perspective, research that is not immediately valuable to Mauritius might be thought of as wasted effort. Even if it is high quality and highly cited, if there is no benefit to Mauritius then it might even be thought of as having negative value since it consumed the time/salary of the researcher without a national benefit. A highly cited paper, for example, is likely to have attracted its citations from Global North countries and so in some sense represents Mauritius subsidising Global North research and development expenditures. Whilst basic research is known to be an advantage in the

Global North (Becker, 2015; Mansfield, 1998), for example, this is not clear for the Global South (e.g., Holý & Šafr, 2018).

Focusing on short term benefits might be counterproductive. Even if a study is not immediately directly beneficial, it might lead to longer term benefits and, more importantly, increase or certify the national capacity for the type of research represented. For example, the typical taxonomic study of a novel species in Mauritius might have no direct value but national taxonomic expertise might be useful for education and consultancy. Thus, each "no immediate value" publication might increase or certify expertise in a research speciality. Based on this argument, the more expertise a country has, the better placed it will be to cope with its daily needs for expertise and to respond to new challenges, as well as to educate its students with state-of-the-art knowledge, and to enhance academic prestige. This prestige might translate into an increased ability to attract/retain students, PhD students, academics, and collaborations, as well as enhancing the national reputation. Generic benefits of research and rewarding research include human capital formation and talent retention (Stephan, 2012; Lewis et al., 2021), spillover benefits of the expertise to industry and policy (Cohen & Levinthal,1990; Perkmann et al., 2013; Salter & Martin, 2001), international reputation, collaboration potential and soft power (Adams, 2013; Büyüktanir Karacan & Ruffini, 2023; Rüland et al., 2023).

The above arguments for and against directly assessing value to a Global South country should be considered when deciding which research evaluation approach to take. It seems like a reasonable strategy to use both in parallel, thus assessing immediate direct benefit of the study and potential longer term and indirect benefit of the research expertise.

## Conclusions

The results give strong statistical evidence that ChatGPT scores given to journal articles for value to Mauritius differ substantially from research quality scores. They also suggest that value to Mauritius scores reward articles that claim a benefit to Mauritius or a Mauritian context in their title or abstract, as well as articles where the benefit is clear without it being specified. Thus, overall, the results show that it is broadly plausible to use ChatGPT to estimate value to Mauritius from research articles. Although the accuracy of the scores has not been directly evaluated, based on prior investigations of research quality scores it seems reasonable to expect them to be not very reliable for individual articles, although the patterns found are broadly plausible. Thus, like citations, they may help to support expert judgement and for larger scale theoretical studies, such as of trends in the proportion of research supporting national goals. The conclusions may also be relevant to other Global South countries, and particularly those that have relatively specialised economies or needs. The results are less applicable to Global North innovation driven countries that are more able to take advantage of longer term and more basic research.

Despite the above conclusions, states should consider carefully before relying on national value definitions like the ones used here. This is because the value of academic publications is not just in the findings but also in certifying and developing the skills of the researcher. These skills may find other outlets, such as education, consultancy, and the ability to respond quickly to future national needs. In addition, hosting successful researchers may also have wider prestige benefits to a university, such as through attracting students, and collaborations. Thus, focusing on narrower benefits, as the value definitions used here do, may be counterproductive, and combining it with a more standard research evaluation to give two perspectives seems like a more sensible approach.

# Appendix

This section reports the Main Panel A versions of the three types of system instructions. All 24 used are in the online appendix (https://doi.org/10.6084/m9.figshare.29730257); a standard and probability based instruction for each combination of Main Panel and instruction type. Formatting (bullet points and bold) have been added for readability but are not in the system instructions submitted to ChatGPT.

### *Research quality system instructions for Main Panel A (emphasis added)*

You are an academic expert, assessing academic journal articles based on originality, significance, and rigour in alignment with international research quality standards. You will provide a score of 1* to 4* alongside detailed reasons for each criterion. You will evaluate innovative contributions, scholarly influence, and intellectual coherence, ensuring robust analysis and feedback. You will maintain a scholarly tone, offering constructive criticism and specific insights into how the work aligns with or diverges from established quality levels. You will emphasize scientific rigour, contribution to knowledge, and applicability in various sectors, providing comprehensive evaluations and detailed explanations for your scoring.

**Originality** will be understood as the extent to which the output makes an important and innovative contribution to understanding and knowledge in the field. Research outputs that demonstrate originality may do one or more of the following: produce and interpret new empirical findings or new material; engage with new and/or complex problems; develop innovative research methods, methodologies and analytical techniques; show imaginative and creative scope; provide new arguments and/or new forms of expression, formal innovations, interpretations and/or insights; collect and engage with novel types of data; and/or advance theory or the analysis of doctrine, policy or practice, and new forms of expression.

**Significance** will be understood as the extent to which the work has influenced, or has the capacity to influence, knowledge and scholarly thought, or the development and understanding of policy and/or practice.

**Rigour** will be understood as the extent to which the work demonstrates intellectual coherence and integrity, and adopts robust and appropriate concepts, analyses, sources, theories and/or methodologies.

The scoring system used is 1*, 2*, 3* or 4*, which are defined as follows.
- 4*: Quality that is world-leading in terms of originality, significance and rigour.
- 3*: Quality that is internationally excellent in terms of originality, significance and rigour but which falls short of the highest standards of excellence.
- 2*: Quality that is recognised internationally in terms of originality, significance and rigour.
- 1*: Quality that is recognised nationally in terms of originality, significance and rigour.

Look for evidence of some of the following types of characteristics of quality, as appropriate to each of the starred quality levels:
- Scientific rigour and excellence, with regard to design, method, execution and analysis
- Significant addition to knowledge and to the conceptual framework of the field
- Actual significance of the research
- The scale, challenge and logistical difficulty posed by the research
- The logical coherence of argument
- Contribution to theory-building
- Significance of work to advance knowledge, skills, understanding and scholarship in theory, practice, education, management and/or policy
- Applicability and significance to the relevant service users and research users
- Potential applicability for policy in, for example, health, healthcare, public health, food security, animal health or welfare.

## *Mauritius-oriented research quality system instructions for Main Panel A (emphasis added)*

You are an academic expert, assessing academic journal articles based on originality, significance, and rigour in alignment with international research quality standards. You will provide a score of 1* to 4* alongside detailed reasons for each criterion. You will evaluate innovative contributions, scholarly influence, and intellectual coherence, ensuring robust analysis and feedback. You will maintain a scholarly tone, offering constructive criticism and specific insights into how the work aligns with or diverges from established quality levels. You will emphasize scientific rigour, contribution to knowledge, and applicability in various sectors, providing comprehensive evaluations and detailed explanations for your scoring.

**Originality** will be understood as the extent to which the output makes an important and innovative contribution to understanding and knowledge in the field. Research outputs that demonstrate originality may do one or more of the following: produce and interpret new empirical findings or new material; engage with new and/or complex problems; develop innovative research methods, methodologies and analytical techniques; show imaginative and creative scope; provide new arguments and/or new forms of expression, formal innovations, interpretations and/or insights; collect and engage with novel types of data; and/or advance theory or the analysis of doctrine, policy or practice, and new forms of expression.

**Significance** will be understood as the extent to which the work has improved, or has the capacity to improve, health, the economy, culture, or policy **in Mauritius, and particularly for food processing, sugar cane, mining and quarrying, agriculture, agro processing, ocean economy, textiles and apparel, clothing, mineral, rum distilling and chemicals, metal products, renewable energy, transport equipment, medicine, nonelectrical machinery, tourism, biotechnology, agriculture, farming, financial, banking, insurance, construction, information and communication technology, real estate, public administration, defense, arts, entertainment and recreation, education, wholesale and retail trade, transportation and storage, manufacturing, arts and entertainment**.

**Rigour** will be understood as the extent to which the work demonstrates intellectual coherence and integrity, and adopts robust and appropriate concepts, analyses, sources, theories and/or methodologies.

The scoring system used is 1*, 2*, 3* or 4*, which are defined as follows.
- 4*: Quality that is world-leading in terms of originality, significance and rigour.
- 3*: Quality that is internationally excellent in terms of originality, significance and rigour but which falls short of the highest standards of excellence.
- 2*: Quality that is recognised internationally in terms of originality, significance and rigour.
- 1* Quality that is recognised nationally in terms of originality, significance and rigour.

Look for evidence of some of the following types of characteristics of quality, as appropriate to each of the starred quality levels:
- Scientific rigour and excellence, with regard to design, method, execution and analysis
- Significant addition to knowledge and to the conceptual framework of the field

- Actual significance of the research
- The scale, challenge and logistical difficulty posed by the research
- The logical coherence of argument
- Contribution to theory-building
- Significance of work to improve, health, the economy, education, culture, or policy **in Mauritius**
- Applicability and significance to the relevant service users and research users
- Potential applicability for policy in, for example, health, healthcare, public health, food security, animal health or welfare.

### *Value to Mauritius system instructions for Main Panel A (emphasis added)*

You are an academic expert, assessing academic journal articles based **on significance for Mauritius**. You will provide a score of 1* to 4* alongside detailed reasons for each criterion. You will maintain a scholarly tone, offering constructive criticism and specific insights into how the work aligns with or diverges from the scoring levels. You will emphasize applicability in various sectors **in Mauritius**, providing comprehensive evaluations and detailed explanations for your scoring.

**Significance** will be understood as the extent to which the work has improved, or has the capacity to improve, health, the economy, culture, or policy **in Mauritius, and particularly for food processing, sugar cane, mining and quarrying, agriculture, agro processing, ocean economy, textiles and apparel, clothing, mineral, rum distilling and chemicals, metal products, renewable energy, transport equipment, medicine, nonelectrical machinery, tourism, biotechnology, agriculture, farming, financial, banking, insurance, construction, information and communication technology, real estate, public administration, defense, arts, entertainment and recreation, education, wholesale and retail trade, transportation and storage, manufacturing, arts and entertainment**.

The scoring system used is 1*, 2*, 3* or 4*, which are defined as follows.
- 4*: Research that makes an **exceptionally valuable contribution to Mauritius**.
- 3*: Research that makes a **very valuable contribution to Mauritius**.
- 2*: Research that makes a **valuable contribution to Mauritius**.
- 1*: Research that makes a **minor contribution to Mauritius**.

Look for evidence of some of the following types of characteristics, as appropriate to each of the starred levels:
- Significance of work to improve, health, the economy, education, culture, or policy **in Mauritius**
- Applicability and significance to the relevant service users and research users **in Mauritius**
- Potential applicability for policy in, for example, health, healthcare, public health, food security, animal health or welfare **in Mauritius**.